\documentclass{article}
\usepackage{microtype}

\usepackage{PRIMEarxiv}

\usepackage[utf8]{inputenc} 
\usepackage[T1]{fontenc}    
\usepackage{hyperref}       
\usepackage{url} 
\usepackage{xcolor}
\usepackage{booktabs}       
\usepackage{amsfonts}       
\usepackage{nicefrac}       
\usepackage{microtype}      
\usepackage{lipsum}
\usepackage{fancyhdr}       
\usepackage{graphicx}       
\graphicspath{{media/}}     
\usepackage{amsmath}
\title{Predicting Turbulence Structure In Street-Canyon Flows using Deep Generative Modeling}

\author{
  Tomek Jaroslawski*, Aakash Patil and Beverley McKeon \\
  Center for Turbulence Reserch \\
  Stanford University \\
  \texttt{*Corresponding author:tomek@stanford.edu}\
}

\begin{document}
\maketitle

\begin{abstract}

\textcolor{black}{The high dimensionality and complex dynamics of turbulent flows in urban street canyons present significant challenges for wind and environmental engineering, particularly in addressing air quality, pollutant dispersion, and extreme wind events. This study introduces a deep learning framework to predict spatio-temporal flow behavior in street canyons with varying geometric configurations and upstream roughness conditions. A convolutional encoder-decoder transformer model, trained on particle image velocimetry (PIV) data from wind tunnel experiments, is employed with autoregressive training to predict flow characteristics. The training dataset contains diverse flow regimes, with a focus on the wall-parallel plane near the canyon roof, a critical region for pollutant exchange between the outer flow and the canyon interior. The model accurately reproduces key flow features, including mean turbulent statistics, two-point correlations, quadrant analysis, and dominant flow structures, while demonstrating strong agreement with experimental data in capturing the temporal evolution of flow dynamics. These findings demonstrate the potential of deep learning models to enhance predictive capabilities for urban canyon flows, offering a pathway toward improved urban design, sustainable environmental management, and more effective pollutant dispersion modeling.}

\end{abstract}


\section{Introduction}

Air quality in urban environments is a pressing contemporary issue, with significant socio-economic implications. The geometric complexity of built areas and the interaction of numerous thermodynamic processes challenge our comprehension of the urban climate. Turbulence plays a fundamental role in the instantaneous dynamics of airflow. Specifically, the atmospheric flow, combined with the complex geometry of the urban canopy, exhibits pronounced multi-scale characteristics, both in space and time. The canyon geometry and upstream roughness are key parameters influencing the flow dynamics, where non-linear amplitude modulations arise between the large-scale structures in the boundary layer and the separated, low-frequency flapping shear layer at the roof level \cite{blackman2019assessment}. Understanding the spatio-temporal structure of these flows is therefore crucial, particularly for analyzing transient phenomena such as accidental pollutant releases or predicting flow states with limited sensor data.

Urban flows, characterised by high Reynolds numbers and complex turbulent behaviour, pose significant challenges for both experimental measurement and numerical simulation. These difficulties have prompted recent data-driven approaches to better understand the physics governing these flows \cite{martinez2023data}. Machine learning has recently emerged as a powerful method to address these challenges \cite{vinuesa2023transformative}. In the context of urban flows, examples include predicting pollution concentrations from mobile field measurements \cite{alas2022pedestrian}, analysing inter-scale turbulent interactions over obstacle arrays \cite{liu2023study}, estimating drag coefficients on buildings using large eddy simulations (LES) \cite{lu2023using}, and developing reduced-order models of flow dynamics \cite{xiao2019reduced}. 

Despite these advances, machine learning algorithms have only been applied to a limited extent for predicting spatial-temporal turbulent dynamics using experimental datasets, particularly those from wind tunnel investigations of urban flows. Developing machine learning frameworks specifically designed for experimental data would enhance our ability to model and predict complex urban flow phenomena, supporting the design and optimisation of urban environments. 

Several studies have focused on spatial and temporal reconstruction, as well as spatial supersampling \cite{schmidt2021machine}. Hybrid deep neural network architectures have been designed to capture the spatial-temporal features of unsteady flows \cite{han2019novel}, and machine learning--based reduced-order models have been proposed for three-dimensional  flows \cite{nakamura2021convolutional}. For instance, a deep learning framework that combines long short-term memory (LSTM) networks and convolutional neural networks (CNNs) has been employed to predict the temporal evolution of turbulent flames \cite{ren2021predictive}. Cremades et al. \cite{cremades2024identifying} used an explainable deep-learning approach with U-net and SHAP (SHapley Additive exPlanations) to study wall-bounded turbulence. Their findings revealed that the most important flow structures for prediction were not always those contributing most to Reynolds shear stress, highlighting potential applications in flow control.

New deep learning architectures, such as transformers, are emerging as powerful tools for temporal problems involving structured and unstructured data. Inspired by convolutional neural networks, transformers use self-attention, a mechanism that evaluates the importance of other data points in the dataset relative to the current one, without relying on recurrence, which is characteristic of models like LSTMs. Autoregression refers to a method where the model predicts future data points based on past predictions, iteratively building sequences over time. These features allow transformers to excel in natural language processing tasks and have made them a strong alternative to traditional recurrent neural networks. Transformers have also been applied in spatio-temporal contexts, such as video analysis. However, they have not yet been utilized for spatio-temporal prediction in experimental flow fields involving turbulent flows in urban street canopies.

In this work, we develop a convolutional encoder-decoder transformer model with autoregressive training to predict spatio-temporal flow behavior from an experimental particle image velocimetry (PIV) dataset of turbulent flows in street canyons. \textcolor{black}{The training dataset spans a range of geometric configurations and upstream roughness conditions, focusing on the wall-parallel plane near the canyon roof—a critical region for pollutant exchange between the canyon interior and the outer flow.} The model's predictive capability is rigorously evaluated through comparisons of mean turbulent statistics, two-point correlations, quadrant events, and modal decomposition results with experimental ground truth data. Furthermore, the model’s ability to accurately capture the time evolution of flow dynamics is assessed, highlighting its capacity to resolve complex spatio-temporal processes. \textcolor{black}{These results show the potential of deep learning frameworks in improving the predictive modeling of turbulent urban flows, contributing to enhanced urban design, sustainable environmental management, and more effective pollutant dispersion modeling.}

\section{Experimental data}
\label{sec:data}
The experimental data, as referenced in \cite{jaroslawski2019spanwise,jaroslawski2020roof}, was acquired from wind tunnel experiments at École Centrale de Nantes, France. These experiments utilized a low-speed boundary-layer wind tunnel with dimensions of 2 meters in width, 2 meters in height, and 24 meters in length, featuring a 5:1 inlet contraction ratio. Figure 1 shows the experimental setup. A simulated suburban atmospheric boundary layer at a 1:200 scale was created at the model location using a combination of three vertical, tapered spires at the inlet, a solid fence 300 mm in height located 1.5 meters downstream of the inlet, and an 18.5-meter fetch of either 2D or 3D roughness elements. The study investigated street canyons with width-to-height aspect ratios of 1 and 3. Stereoscopic particle image velocimetry (PIV) measurements were conducted horizontally ($x-z$ plane) at a height of 0.9$h$ ± 0.05$h$. The PIV setup, positioned beneath the wind tunnel floor, utilized a Litron double cavity laser and DANTEC Dynamic Studio software to produce vector fields with a spatial resolution of 1.6 mm and a sampling frequency of 7 Hz. Velocity vector fields were computed using an iterative cross-correlation analysis with a window size of 64 × 64 pixels and an interrogation window of 32 × 32 pixels, overlapped by 50\%, and a pulse interval of 500 $\mu$s. Measurement uncertainties for mean velocity, standard deviation, and turbulent shear stress were estimated at 0.9\%, 1.4\%, and 3.9\%, respectively, based on 2551 independent samples from 10,000 velocity field recordings. The 10,000 snapshots correspond to approximately 170,000 eddy turnovers, where the eddy turnover time is defined as $T = h / U_{e}$. The freestream velocity of $U_{e}$ = 5.9 ms$^{-1}$, measured using a pitot-static tube, remained constant across experiments, yielding a Reynolds number of $1.9 \times 10^4$ based on this speed and canyon height, $h$. \textcolor{black}{Real-world street canyons typically have Reynolds numbers (\( Re_{h} \)) ranging from \( 10^5 \) to \( 10^6 \), influenced by building height, street width, and wind speed. In contrast, laboratory and wind tunnel studies usually operate at lower \( Re_{h} \) values (\( 10^3 \) to \( 10^5 \)), with Reynolds number independence often assumed at \( Re_h \approx 3 \times 10^4 \) \cite{castro1977flow}.}

To address the challenge of training the model on data significantly different from the prediction target, we train our model on a dataset with an aspect ratio \(AR=3\), while the prediction dataset has a different aspect ratio of \(AR=1\). This difference in aspect ratios results in datasets of varying sizes. Additionally, experimental PIV data often exhibit poor quality near the walls, leading to incomplete or unreliable measurements. As a result, experimental datasets are not always uniform in size due to the exclusion of low-quality boundary data. To standardise the input dimensions across datasets, we employ zero-padding, as depicted schematically in Fig. \ref{fig:2}. This approach ensures compatibility between the training and prediction datasets while accounting for the inherent variability in experimental data.

\begin{figure}[h!]
    \centering
\includegraphics[width=0.8\textwidth]{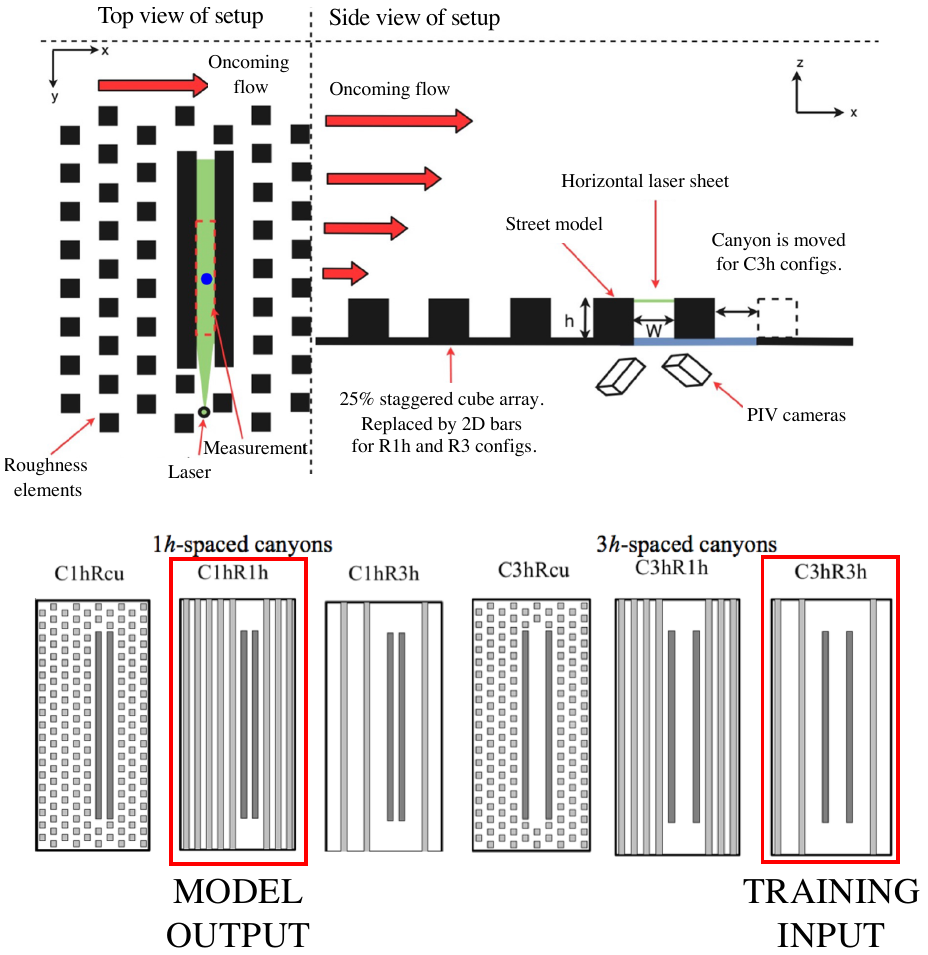} 
    \caption{(Experimental setup (top) and configurations (bottom).}
    \label{fig:1}
\end{figure}

\begin{figure}[h!]
    \centering
\includegraphics[width=1\textwidth]{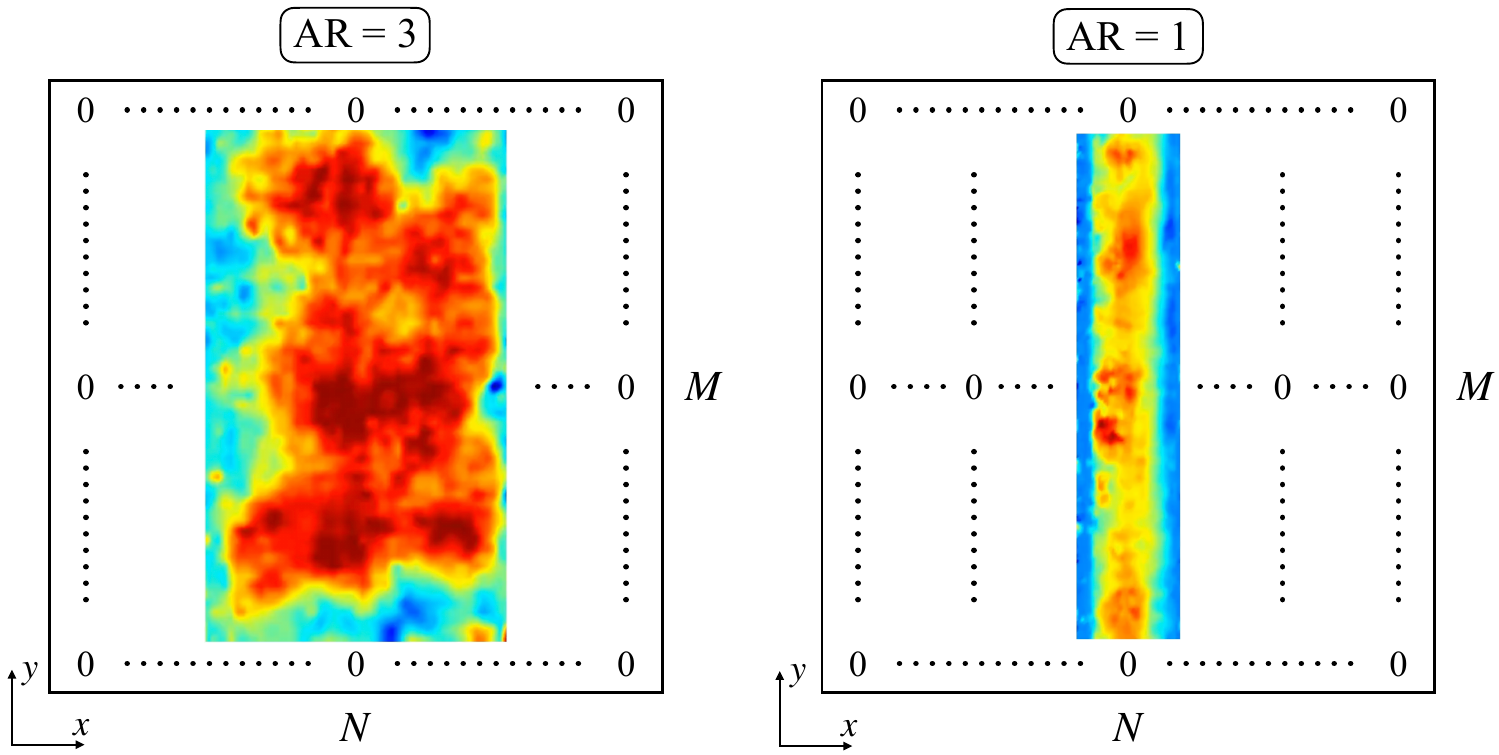} 
    \caption{Streamwise velocity field snapshots of experimental data for both wide ($AR=3$) and narrow ($AR=1$) street canyons, with zero-padding applied to standardise the data size to $N \times M$. This preprocessing step ensures consistent dimensionality across different canyon geometries.}
    \label{fig:2}
\end{figure}

\section{Deep learning framework}

\begin{figure}[h!]
    \centering
\includegraphics[width=1\textwidth]{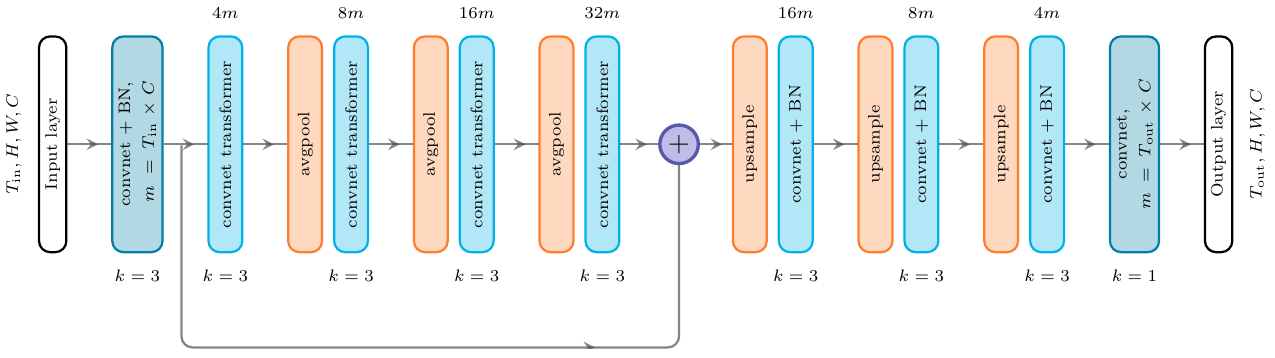} 
    \caption{Model architecture of the convolutional encoder-decoder transformer to process low and high level features. The canonical four-stage design is utilized in  addition to the convolutional transformer blocks or layers. $H,W$ are the input resolutions for each snapshot in $T_{\text{in}}$ sequence and $T_{\text{out}}$ sequence, $k$ is the kernel size, and $m$, the number of filters. }
    \label{fig:3}
\end{figure}

Deep Generative Models, specifically the fusion of transformers with convolutional encoder-decoder architectures, offer state-of-the-art performance for modeling complex turbulent flow dynamics. This hybrid approach combines the transformer's ability to capture long-range dependencies with the convolutional neural network's ability in extracting hierarchical spatial features. The self-attention mechanism, when integrated within convolutional layers, enhances the model's capacity to identify and prioritize important features while suppressing less significant features. Figure \ref{fig:3} illustrates our integrated generative framework, which we employ for predicting spatio-temporal evolution of turbulent structures.

In our generative modeling of street-canyon turbulence, we process a time-series comprising $N$ sequential flow field snapshots $\left[ x_{t},x_{t+\Delta t}, ...., x_{t+(N-1)\Delta t} \right]$, with the objective of generating physically consistent flow fields $M$ steps ahead in time. The input $X$ to our deep generative model consists of a sequence $T_{\text{in}}$ of flow field snapshots $\left[ x_{t},x_{t+\Delta t}, ...., x_{t+(N-1)\Delta t} \right] $, while the generated output $Y$ comprises a sequence $T_{\text{out}}$ of predicted flow fields $\left[ x_{t+N\Delta t}, ..., x_{t+N+(M-1)\Delta t} \right]$. Each snapshot \( x_t \) contains the streamwise, vertical, and spanwise velocity components.

The generative encoder processes input flow fields with \(H\times W\) spatial resolutions, learning a latent representation that captures both large-scale coherent structures and small-scale turbulent features. The decoder then reconstructs this compressed representation into physically consistent output flow fields through progressive up-sampling and convolutions. The weight matrices of both components enable the model to capture and learn the multi-scale turbulent dynamics present in street canyon flows. At \(x_{t+\Delta t}\), the decoder reconstructs the flow field while preserving important turbulent statistics.

During the encoding process, input flow sequences are concatenated channel-wise and processed through successive convolutional operations. The intermediate feature maps \(\mathbb{F}\in \mathbb{R}^{C\times H\times W}\) with \(C\) channels undergo self-attention in the convolutional transformer layer, which incorporates both spatial information and temporal positioning of the flow sequence. This layer employs a \(3 \times 3\) kernel while maintaining the learned convolutional features, enabling the model to capture both local and non-local turbulent interactions characteristic of street-canyon flows. The convolutional transformer layer facilitates the capture of both spatial and temporal non-locality. \textcolor{black}{The convolutional component effectively captures spatial non-local interactions, while the self-attention mechanism enables the modeling of temporal non-local dependencies.}

The generative framework is trained in an autoregressive manner to ensure temporal consistency in the predicted flow fields. In the context of turbulent flows, autoregressive prediction is particularly important as it allows the model to maintain physical correlations across time steps. For a trained generative model $\mathbb{M}$ shown in Figure \ref{fig:3}, multi-step prediction of flow quantity $X_{t}$ follows an auto-regressive sequence, where each predicted state serves as input for the subsequent prediction:
\begin{equation}
	\left\{
	\begin{split}
		\widehat{X}_{t+\Delta t} 	&= \mathbb{M}(X_{t}), \\
		\widehat{X}_{t+2\Delta t} 	&= \mathbb{M}(\widehat{X}_{t+\Delta t}), \\
							&\,\, ... \\
		\widehat{X}_{t+(n-1)\Delta t} 	&= \mathbb{M}(\widehat{X}_{t+(n-2)\Delta t}), \\
	\end{split}
	\right.
\end{equation}
\noindent where $t$ is the time step and $X \in \mathbb{R}^{C \times H \times W}$ is the input tensor snapshot at instant $t$. In the following, the autoregressive training sequence length is set equal to two in order to limit the computational cost. 
We train the model by employing the Adam optimizer \cite{kingma2014adam} to iteratively minimize the total equi-weighted mean squared error (MSE) loss defined by:
\begin{equation}
	\begin{split}
		\\\mathcal{L} = \frac{1}{n_s} \biggr[ &\sum_{i=1}^{n_s} \left( \left(X_{t+\Delta t} \right)^{i}  - \left(\widehat{X}_{t+\Delta t} \right)^{i} \right)^2 +\, \cdots	\\
								+& \sum_{i=1}^{n_s} \left( \left(X_{t+2\Delta t} \right)^{i}  - \left(\widehat{X}_{t+2\Delta t}\right)^{i} \right)^2 + \, \cdots \\
								+&  \sum_{i=1}^{n_s} \left( \left(X_{t+(n-1)\Delta t} \right)^{i} - \left(\widehat{X}_{t+(n-1)\Delta t}\right)^{i} \right)^2 \biggr].
	\end{split}
\label{lossRelation}
\end{equation}

\textcolor{black}{The loss function actively minimizes the average squared prediction error by summing over all training samples in the dataset (\( n_s \)) and the time steps in the prediction sequence (\( n \)). The autoregressive nature of the model ensures that each prediction depends on the previous one, reinforcing temporal consistency throughout the sequence.}

The neural network utilizes the ReLU activation function, chosen for its effectiveness in handling the nonlinear nature of turbulent flows while maintaining stable weight updates during training \cite{nair2010rectified}. The training process involves the iterative presentation of the flow field dataset, with random shuffling of input-output data pairs over time to ensure robust learning of turbulent statistics. An early stopping criterion, combined with learning rate reduction after 100 epochs of stagnant progress, prevents overfitting to specific flow patterns. The deep generative architecture is implemented using TensorFlow \cite{abadi2016tensorflow}, with training executed on an Nvidia RTX A4500 GPU to handle the computational demands of turbulent flow prediction.

In this study, we utilize PIV measurements of a street canyon with width 3$h$ and upstream roughness elements (2D bars) spaced at 3$h$ intervals for training. This configuration, representative of the wake-interference flow regime \cite{oke1988street}, features complex interactions between adjacent roughness element wakes. The generative model is trained on 10,000 PIV snapshots and subsequently evaluated in the skimming flow regime (canyon spacing of 1$h$), where the flow dynamics fundamentally differ due to the formation of stable recirculation zones above the closely packed elements.

\section{Results and discussion}

This section presents a comparison of the results, focusing on mean and two-point turbulent statistics, quadrant analysis, and proper orthogonal decomposition (POD) between the PIV reference data and the model predictions for the C1hR1h configuration. \textcolor{black}{For completeness and as a form of validation, we also evaluate the performance of the model in the training data set (C3hR3h configuration) by analyzing the convergence of statistics and turbulent quantities, provided in the Appendix \ref{appen}.}

\subsection{Convergence of turbulent statistics}

\begin{figure}[h!]
    \centering
\includegraphics[width=0.95\textwidth]{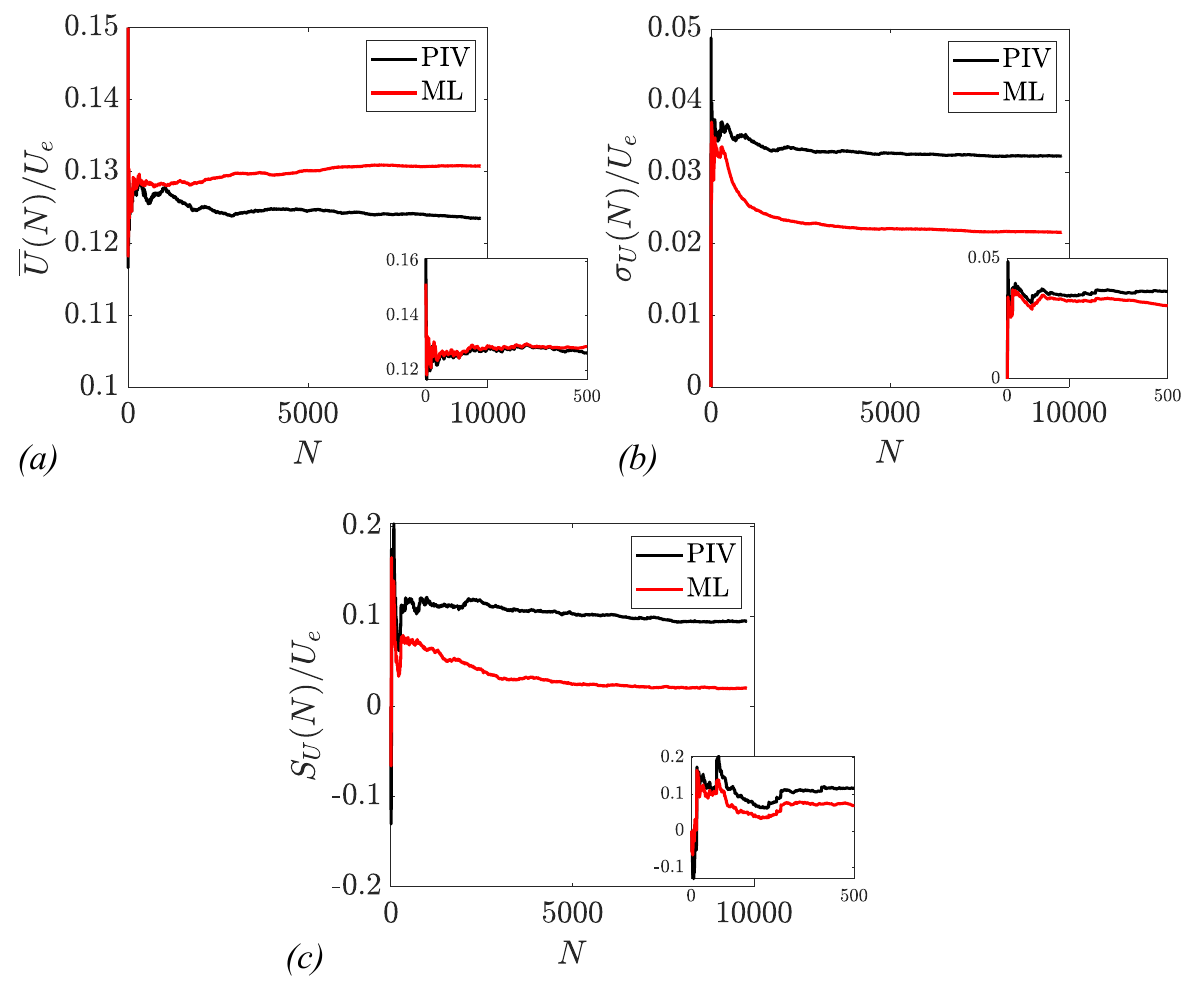} 
    \caption{Convergence of the mean velocity, standard deviation, and skewness at a single point \((x=0, y=0)\) for both the ML model and PIV data.}
    \label{fig:conv}
\end{figure}

In Fig. \ref{fig:conv}, we show the convergence of the mean, standard deviation, and skewness of the streamwise velocity fluctuations at the reference centre point and roof level of the canyon ($x=0$, $y=0$, $z=0.9h$) for the test case C1hR1h. We acquired the experimental PIV data over a sufficiently long duration to ensure statistical convergence. The figure clearly demonstrates convergence, indicating that the model was trained on a substantial volume of data.

From Fig. \ref{fig:conv}, it is evident that the model achieves good performance in reproducing the mean and standard deviation at early times, particularly within the first 500 snapshots. However, at longer time horizons, the model converges to values lower than those observed in the experimental results. 

The model was initialised using the first four snapshots and then allowed to evolve independently. The discrepancies at longer time horizons likely result from cumulative errors in the model, which grow over time and eventually lead to divergence. This issue may stem from the dataset’s lack of temporal resolution, forcing the model to rely primarily on spatial information to infer flow dynamics. Attempts to improve performance by varying the training parameters yielded limited success, suggesting that this limitation might be inherent to the available data. Nevertheless, the transformer model demonstrates strong utilisation of spatial information up to a certain point.

Despite these limitations, the dataset shows agreement between the ML model and experimental results for approximately 8500 eddy turnover times, representing a significant amount of data. For context, the model required less than a minute to generate these data on a single Nvidia RTX A4500 GPU. For the remainder of this paper, we base all analyses on the first 500 snapshots, where the model aligns well with the experimental data.

\subsection{Turbulent statistics}

\begin{figure}[h!]
    \centering
\includegraphics[width=0.9\textwidth]{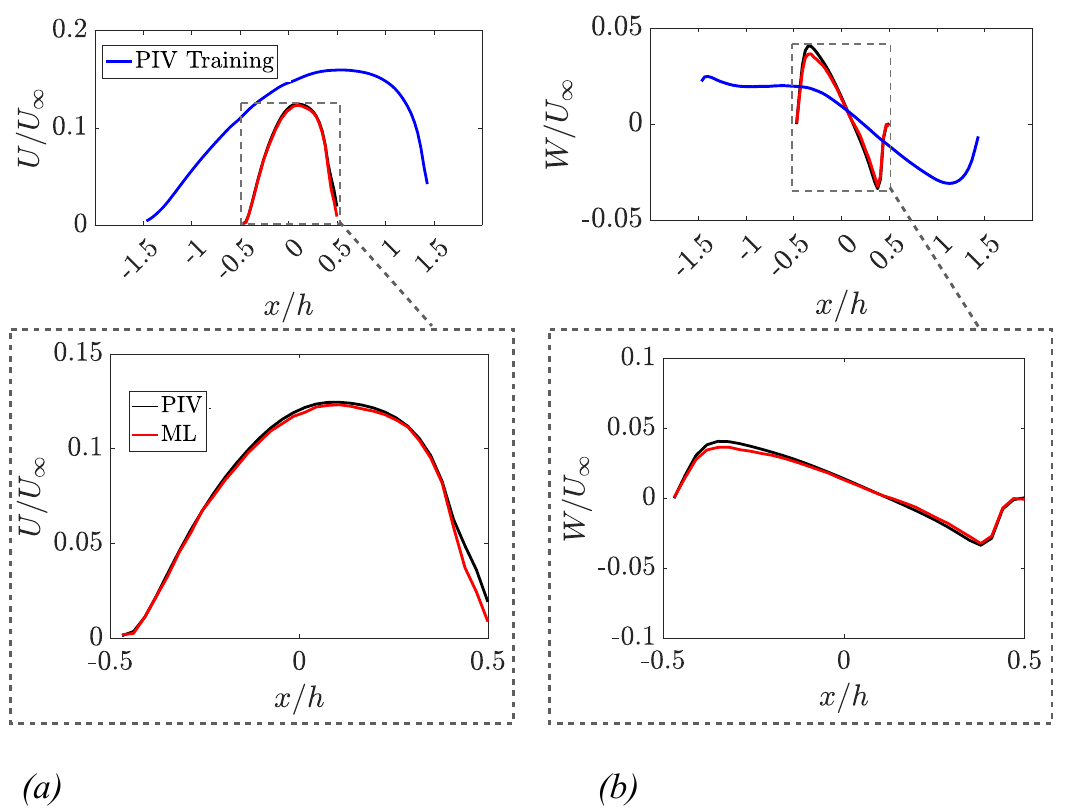} 
    \caption{Time- and spatially-averaged mean velocity profiles for (a) streamwise velocity and (b) vertical velocity. The blue line represents the training configuration (Ch3R3h), while the red and black lines denote the test case (C1hC1h).}
    \label{fig:mean}
\end{figure}

Figure \ref{fig:mean} presents a comparison between the model's predictions and the experimental data for the time- and spanwise-averaged streamwise and vertical velocity profiles as functions of the \(x\)-direction. The mean flow profiles in the training data, presented in blue in Fig. \ref{fig:mean}, differ significantly from those in the test case. This difference arises because the canyon width-to-height ratio is reduced to 1 in the test case compared to 3 in the training data. Additionally, the test case is in a skimming flow regime, which differs from the wake interaction flow regime present in the training data \cite{oke1988street}. Notably, the \(W\) velocity component in the training dataset, as shown in Fig. \ref{fig:mean}b, exhibits a larger and asymmetric recirculation region skewed towards the windward side of the canyon. This behaviour contrasts with the profile observed in the test case. These significant differences between the training and test datasets allow us to critically assess the model's predictive capabilities. Despite these variations, the mean flow and vertical velocity profiles presented in Fig. \ref{fig:mean} show strong agreement between the predictions of the model and the experimental data for the mean flow.

\begin{figure}[h!]
    \centering
\includegraphics[width=1\textwidth]{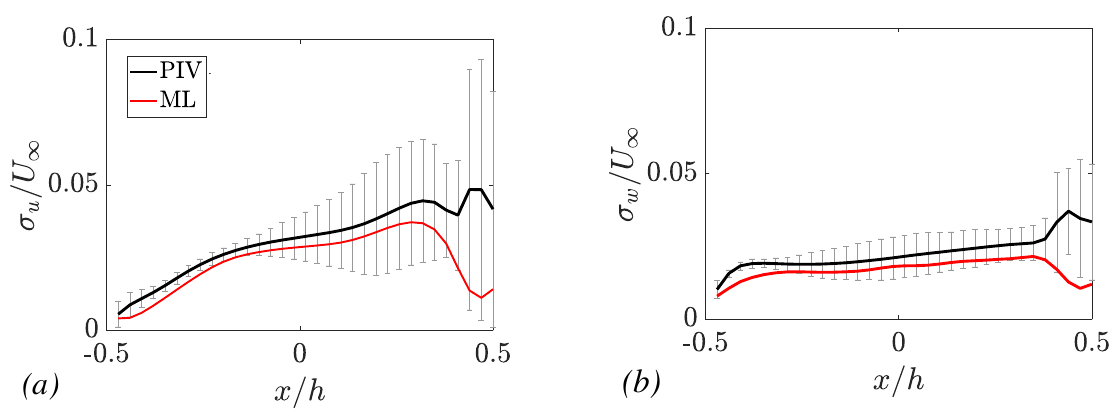} 
    \caption{Time- and spatially-averaged standard deviation profiles of (a) streamwise velocity and (b) vertical velocity components for the C1hR1h test case.}
    \label{fig:std}
\end{figure}

Figure \ref{fig:std} presents the profiles of the spanwise-averaged standard deviation at the streamwise roof level for both the streamwise (\( \sigma_{u} \)) and vertical (\( \sigma_{w} \)) velocity components. The error bars indicate the spanwise variation of \( \sigma_{u} \) and \( \sigma_{w} \), calculated from the experimental data. Overall, the spatial evolution of these profiles shows good agreement with the experimental results.  

The model's predictions generally exhibit lower values for \( \sigma_{u} \) and \( \sigma_{w} \). Despite this, the predicted values mostly fall within the spanwise variation of the experimental statistics, indicating that the model captures the broader trends reasonably well. However, the discrepancy between the ML predictions and the experimental data becomes more pronounced downstream in the canyon, especially near the windward side. This region of the canyon is physically characterised by increased flow intermittency and higher levels of turbulence. These phenomena arise from interactions between the oncoming wind and the building surfaces, as well as from the formation of complex recirculation patterns. Consequently, the flow in this region is more chaotic compared to the leeward side. The poorer performance of the ML model in the windward region may be attributed to several factors. One key limitation could be insufficient resolution in the training data, which may fail to capture the fine-scale turbulent structures that dominate this part of the flow. Furthermore, the model's reliance on spatial information, rather than temporal dynamics, may hinder its ability to accurately predict highly intermittent or transient flow behaviors.  
  
To further assess the model's capacity in predicting turbulence structure, we conducted two-point spatial correlation analysis on the streamwise velocity fluctuations, represented as \(R_{uu}\). Two-point spatial fluctuating velocity correlations offer important information regarding the structure of the flow field that single-point measurements are unable to provide. A two-point spatial correlation was conducted using the middle of the street canyon as the reference point. The two-point correlation coefficient was computed using

\begin{equation}
R_{uu} = \frac{\overline{u'(x_{ref},y_{ref})u'(x,y)}}{\sqrt{\overline{u'(x_{ref},y_{ref})}}\sqrt{\overline{u'(x,y)}}}.
\end{equation}

\begin{figure}[h!]
    \centering
\includegraphics[width=1\textwidth]{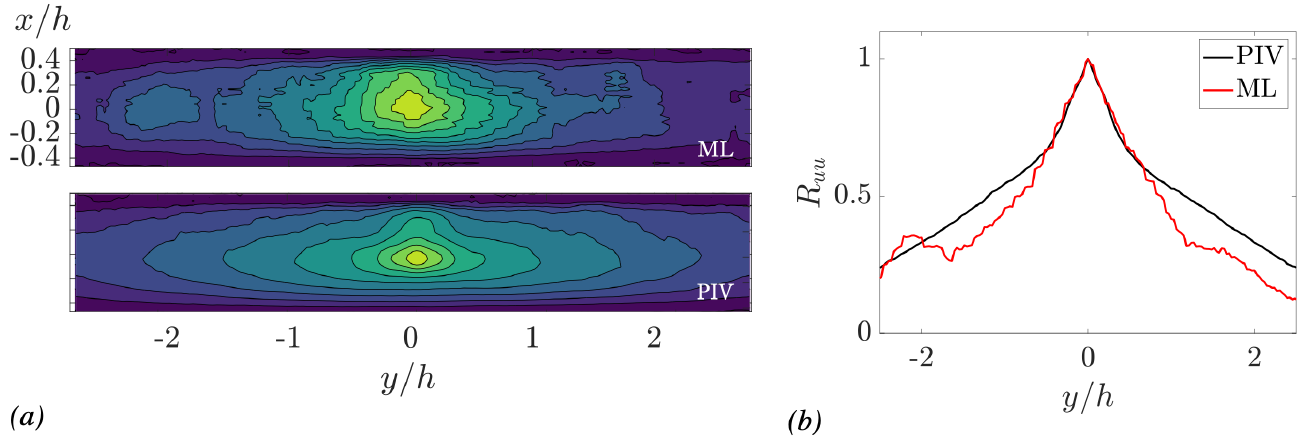} 
    \caption{(a) Two-point correlation fields, $R_{uu}$, of the ML model and PIV dataset. (b) Spanwise slice at $x=0$ of the $R_{uu}$ field.}
    \label{fig:corr}
\end{figure}

Figure \ref{fig:corr}a shows contour fields of $R_{uu}$ for both the ML model and the PIV data. The ML model effectively captures the general spatial structure, particularly near the reference point $x_{ref}=0, y_{ref}=0$. In Fig. \ref{fig:corr}b, a spanwise slice is presented at the streamwise position of $x=0$. This further highlights the model's capability to predict fluctuations' decorrelation at smaller spatial lags, indicating its potential in forecasting the spanwise turbulence structure. At greater spatial lags, the correlation decreases more rapidly than in the experimental data, possibly due to energy dissipation by the model. \textcolor{black}{Furthermore, the \(x\)-variation of \(R_{uu}\) approaching the windward end has a lower magnitude than observed in the experimental data, which aligns with the lower fluctuations in the standard deviation profiles predicted by the model in Fig. \ref{fig:std}.} It is worth noting that the noise observed in the ML results may stem from the utilization of 500 snapshots for computations, whereas the experimental dataset incorporated 10,000 snapshots.

\subsection{Temporal dynamics of turbulent structures}

To investigate the temporal evolution of relevant flow structures, we track Q events over time, computed using classical quadrant analysis. This method identifies regions with statistically significant Reynolds stress magnitudes and classifies turbulent events based on the direction of fluctuating velocities. According to quadrant analysis \cite{wallace2016quadrant}, four types of events are defined: Q1 (Outward Interaction), where high streamwise velocity structures (\( u' > 0, w' > 0 \)) move from the wall to the bulk flow; Q2 (Ejection), where low streamwise velocity structures (\( u' < 0, w' > 0 \)) move from the wall to the bulk flow; Q3 (Inward Interaction), where low streamwise velocity structures (\( u' < 0, w' < 0 \)) move from the bulk flow to the wall; and Q4 (Sweep), where high streamwise velocity structures (\( u' > 0, w' < 0 \)) move from the bulk flow to the wall. By tracking the occurrence of these quadrant events over time, we can assess the model’s ability to predict turbulent flow structures and capture the evolution of coherent structures near the roof level of the canyon.

\begin{figure}[h!]
    \centering
\includegraphics[width=0.7\textwidth]{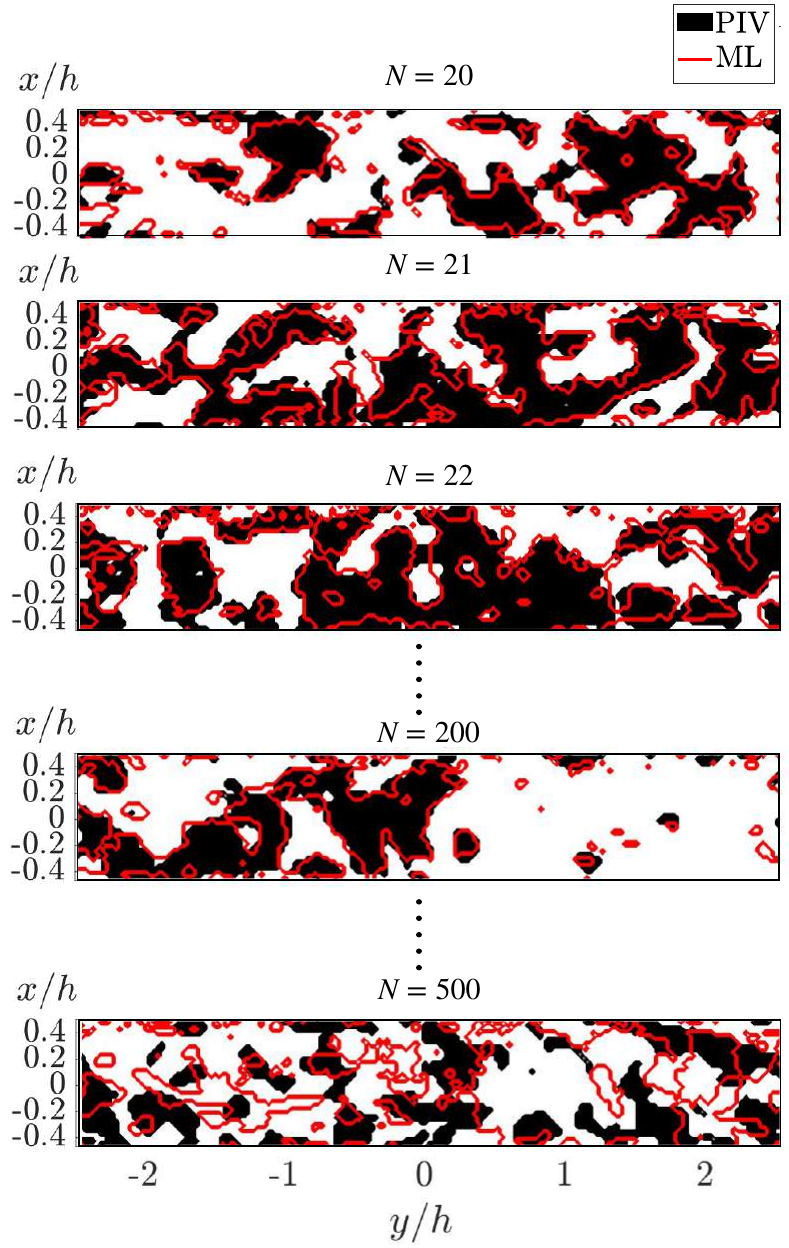} 
    \caption{Temporal evolution of Q2 (ejection) fields from experimental data and the ML model. The black region represents areas where the condition \( u' < 0, w' > 0 \) is satisfied in the experimental data, while the red contour shows the corresponding region predicted by the ML model.}
    \label{fig:q2time}
\end{figure}

In Fig. \ref{fig:q2time}, we present the temporal evolution of Q2 (ejection) events, comparing predictions from the ML model with experimental data for selected snapshots up to \(N = 500\), at which point the model begins to deteriorate. \textcolor{black}{For brevity, only the Q2 events are presented; however, these events are particularly relevant for understanding phenomena such as contaminant exit from the canyon.} The model successfully predicts the locations of ejections, capturing the general large-scale structure of the flow. However, small-scale structures are not fully resolved, likely due to the model filtering out small-scale energy, as discussed earlier. Despite this limitation, the model demonstrates impressive temporal prediction capabilities, accurately forecasting quadrant events well into the temporal evolution. This performance is noteworthy, given that the model is initialized with only the first four snapshots but can predict the flow dynamics over a significant time period.

To approach the quadrant analysis from a more statistical perspective, a hole analysis is applied to measure the contribution of each quadrant to \( u'w' \), by further limiting the data to values above a certain amplitude threshold. The threshold level, \( T_H \), is determined as a multiple (\( H \)) of the root-mean-square (r.m.s.) stress: \( T_H = H(u'w') \).

\begin{figure}[h!]
    \centering
\includegraphics[width=1\textwidth]{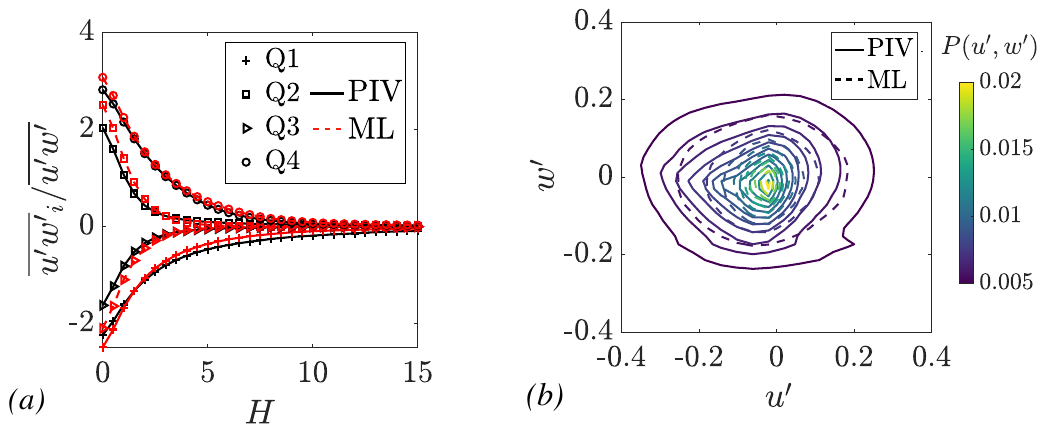} 
    \caption{(a) Q1-Q4 contributions to the total shear stress versus the hole size, $H$ (b) Joint-Probability Function, $P(u',w')$}
    \label{fig:hole}
\end{figure}

The results of the hole analysis are shown in Figure \ref{fig:hole}a, illustrating the contribution of each quadrant to the total shear stress as a function of hole size, $H$. The ML dataset exhibits agreement with the experimental outcomes, a conclusion further supported by examining the probability density function of the streamwise and vertical fluctuations, \(P(u', w')\), shown in Figure \ref{fig:hole}b. Nonetheless, at lower threshold levels (\(H\)), opportunities for enhancement of the model emerge, highlighted by a slightly less precise overlap of the ML and PIV data. This divergence is, at least in part, presumably due to the truncated dataset size within the ML analysis vis-à-vis the exhaustive experimental dataset, indicating that future model refinement should focus on enhancing the model's ability to generate more snapshots.

\subsection{Proper Orthogonal Decomposition}

\textcolor{black}{We apply a snapshot Proper Orthogonal Decomposition (POD) analysis \cite{sirovich1987turbulence} to the fluctuating velocity fields from both experimental and ML-generated datasets. POD identifies and quantifies the dominant flow structures, allowing us to analyze the distribution of kinetic energy across different motion scales. While researchers commonly use POD for higher-order analysis of experimental data, the literature lacks studies that quantify and compare machine learning model results in this way. POD is also a widely used tool for low-order modeling, as it captures the most energetic flow features, which can be reduced to simplified models.} By applying this approach, we evaluate the model’s ability to capture and reproduce the most energetically relevant flow structures, which are crucial for understanding turbulence dynamics in complex flows, such as those in street canyons.

\begin{figure}[h!]
    \centering
\includegraphics[width=1\textwidth]{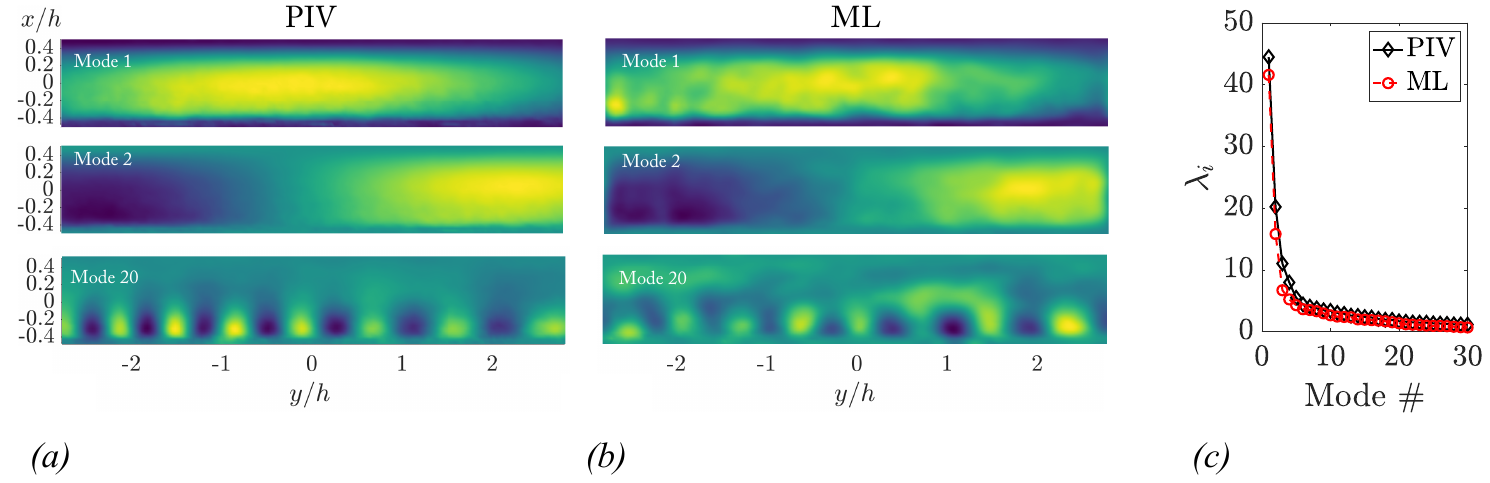} 
    \caption{POD modes 1, 2 and 20 of the PIV (a) and ML (b) data. (c) First 30 eigenvalues.}
    \label{fig:pod}
\end{figure}

In Figure \ref{fig:pod}a,b, we present the spatial structures of POD modes 1, 2, and Mode 20, a higher-order mode, to assess how well the model captures small-scale energy. The first two modes, associated with large-scale structures linked to the low-frequency dynamics of the separated shear layer, show strong structural agreement between the experimental and ML-derived POD modes. This indicates that the model can effectively capture the major flow features that dominate energy distribution. As we move to higher-order modes, such as Mode 20, we observe a reduction in spanwise wavelength and periodicity along the canyon’s span, which corresponds to smaller structures. Notably, the ML model performs well in capturing these fine-scale dynamics in Mode 20, demonstrating that it can replicate the organization of smaller-scale structures within the shear layer at the canyon’s roof level. However, although the general small-scale dynamics are captured, some finer details are lacking. This is corroborated by the quadrant analysis, where certain small-scale features are not fully captured. 

In Figure \ref{fig:pod}c, we show the eigenvalues for the first 30 modes. While the general trend of the eigenvalues is consistent between the experimental and ML data, the eigenvalues of the ML model are systematically smaller across all modes. This suggests a filtering of certain flow features in the ML model. These discrepancies could arise from the limited number of snapshots in the training dataset compared to the experimental reference, which may restrict the model’s ability to fully capture fine-scale dynamics. Despite these differences, the overall agreement suggests that the model performs well in capturing the essential flow dynamics. The POD analysis, therefore, serves as an effective tool for evaluating the fidelity of the ML model in replicating both large- and small-scale flow structures, offering deeper insight into its predictive capabilities.

\section{Conclusions}

In this study, a machine learning framework integrating a convolutional encoder-decoder transformer with autoregressive training was employed to predict spatio-temporal dynamics within a street canyon. The model was trained using wind tunnel PIV measurements collected at the roof level of the canyon in the 
$x-z$ plane. The training dataset configuration comprised a street canyon with a width-to-height ratio of 3, with an upstream roughness fetch consisting of 2D transverse bars, designed to induce a wake interference flow regime. To assess the model's predictive capabilities, we evaluated its performance on a distinct configuration with a reduced width-to-height ratio of 1, where the upstream roughness comprised 2D transverse bars arranged to generate a skimming flow regime.

The results of this study show the effectiveness of the convolutional encoder-decoder transformer with autoregressive training for predicting spatio-temporal dynamics of street canyon flows. The model demonstrated robust replication of the mean flow and the standard deviation of velocity fluctuations, with particularly accurate predictions within the first 500 snapshots. Additionally, the model was able to predict large-scale flow structures, including quadrant events, with a high degree of agreement with experimental data.

However, the model exhibited limitations in capturing fine-scale turbulent structures and transient behaviors, especially in regions with pronounced flow intermittency, such as the windward side of the canyon. Despite these challenges, the model performed well over approximately 8500 eddy turnover times, a sufficiently extensive dataset to enable higher-order analyses such as POD. The POD analysis revealed that while the model accurately reproduced large-scale flow features, it qualitatively captured small-scale dynamics, particularly in the higher-order POD modes.

\textcolor{black}{The proposed framework can be applied to a variety of flow regimes and canyon types, particularly in urban environments with diverse geometries. Its adaptability allows it to capture flow behavior in dense city centers with varying building configurations, street widths, and terrain features. From a practical standpoint, the framework serves as a predictive tool for estimating flow characteristics from limited data or sparse temporal snapshots, making it valuable for early-stage design.}

\textcolor{black}{In urban contexts, the framework can optimize building layouts, improve air quality, and mitigate extreme wind events. By extending the framework to higher Reynolds number flows or more complex geometries, it could provide valuable insights into flow dynamics in diverse urban landscapes. To apply this framework to real-world urban environments, high-quality spatio-temporal data from various street canyon configurations is needed for model training. Wind tunnel experiments or field measurements from existing urban canyons would provide the necessary data, enabling predictions for new or sparsely monitored environments.}

\section{Acknowledgments}

\noindent Partial support of ONR under grant N00014-22-1-2150 is gratefully acknowledged. The data was obtained by T.M. Jaroslawski  during his stay at École Centrale de Nantes, under the supervision of L. Perret and E. Savory.

\newpage

\appendix

\renewcommand{\thefigure}{A\arabic{figure}}
\setcounter{figure}{0} 

\section{\textcolor{black}{Appendix: Model predictions based on the training dataset}}
\label{appen}

In this appendix, we evaluate the model's performance on its training data (C3hR3h) by analyzing the convergence of the turbulent statistics. The mean and standard deviation profiles, averaged both spatially and temporally, are also computed. 

\subsection{Convergence of turbulent statistics}

\begin{figure}[h!]
    \centering
    \includegraphics[width=1\textwidth]{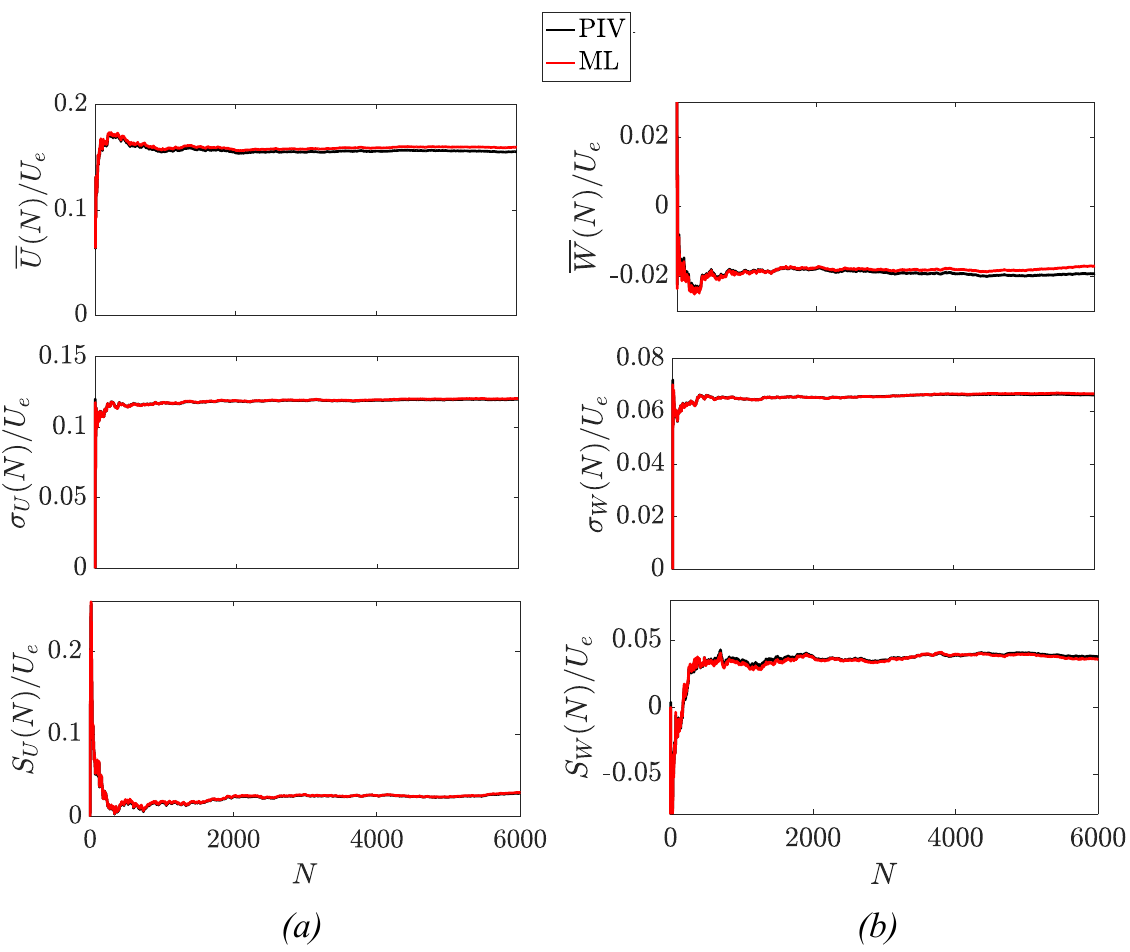}
    \caption{Convergence of the mean velocity, standard deviation, and skewness at a single point \((x=0, y=0)\) for (a) the streamwise and (b) the vertical velocity components, comparing the ML model and PIV data for the training dataset (C3hR3h).}
    \label{fig:appendix1}
\end{figure}

In Fig. \ref{fig:appendix1}, we present the convergence of the mean, standard deviation, and skewness of the streamwise velocity fluctuations at the reference center point and roof level of the canyon \((x=0, y=0, z=0.9h)\) for the training dataset (C3hR3h). Similar to the test data shown in Fig. \ref{fig:conv}, the model was initialized with the first four snapshots before being allowed to evolve independently. For the evaluated time duration, there is excellent agreement in the statistics for both the streamwise (\(U\)) and vertical (\(W\)) velocity components. Notably, the temporal evolution is captured for a significantly longer duration compared to the results in Fig. \ref{fig:conv}, as expected given that the model was trained on this dataset. 

\subsection{Turbulent statistics}

\begin{figure}[h!]
    \centering
    \includegraphics[width=1\textwidth]{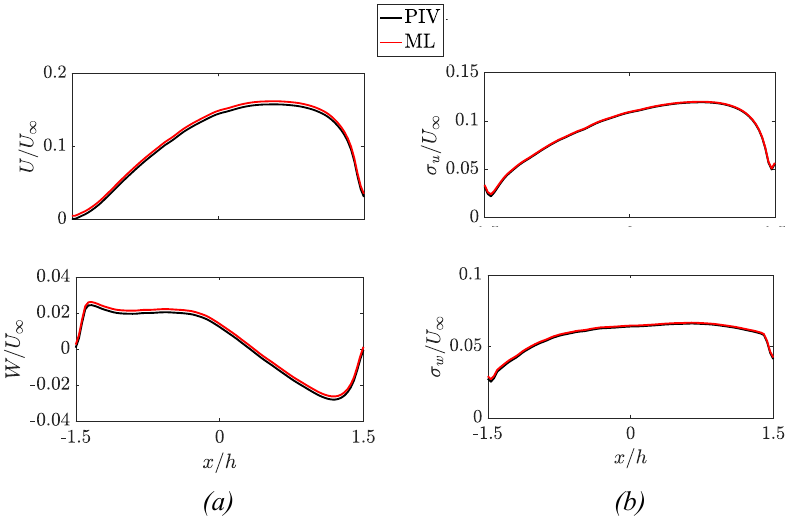}
    \caption{Time- and spatially-averaged profiles for the C3hR3h training data: (a) mean streamwise and vertical velocity profiles, and (b) standard deviation of the streamwise and vertical velocity components.}
    \label{fig:appendix2}
\end{figure}

Here, we present the spatially and temporally averaged mean flow statistics, as in the main text. Figure \ref{fig:appendix2}a compares the time- and spanwise-averaged streamwise and vertical velocity profiles as functions of the \(x\)-direction, contrasting the model's predictions with the experimental data. The results indicate strong agreement between the model and experimental data for both the streamwise and vertical mean velocity profiles.

In Fig. \ref{fig:appendix2}b, we show the profiles of the spanwise-averaged standard deviation at the streamwise roof level for the streamwise (\(\sigma_{u}\)) and vertical (\(\sigma_{w}\)) velocity components. Notably, the agreement between the model predictions and experimental data for \(\sigma_{u}\) and \(\sigma_{w}\) is excellent and surpasses the agreement observed for the test case in Fig. \ref{fig:std}. This improvement is expected, as the model was trained on this dataset, but it also highlights the potential for enhanced predictive performance when the model is exposed to larger and more diverse datasets.

\bibliographystyle{unsrt}  
\bibliography{references}

\end{document}